\documentclass[traditabstract]{aa}
\usepackage{graphicx}
\usepackage{txfonts}
\usepackage{natbib}
\usepackage{soul}
\usepackage{longtable}
\bibpunct{(}{)}{;}{a}{}{,} % to follow the A&A style

\def\solar {\ifmmode_{\mathord\odot} \else $_{\mathord\odot}$ \fi}% _solar
\def\jup {\ifmmode_{\mathrm{Jup}} \else $_{\mathrm{Jup}}$ \fi}
\def\Msol {\ifmmode {\,\mathrm{M}\solar} \else \,M\solar \fi}     % solar mass
\def\Rsol {\ifmmode {\,\mathrm{R}\solar} \else R\solar \fi}     % solar radius
\def\Lsol {\ifmmode {\,\mathrm{L}\solar} \else L\solar \fi}     % solar radius
\def\Mjup {\ifmmode {\,\mathrm{M}\jup} \else M\jup \fi}
\def\Mearth {\ifmmode {\,\mathrm{M}\earth} \else M\earth \fi}

\def\mps {\ifmmode {\,\mathrm{m\,s^{-1}}} \else $\mathrm{m\,s^{-1}}$
\fi}     %meter per sec

\begin{document}
        \title{The HARPS search for southern extra-solar  
        planets\thanks{Based on observations made with the HARPS
        instrument on the ESO 3.6 m telescope under the GTO and LP 
        programs 072.C-0488 and 183.C-0437 at Cerro La Silla (Chile).
        Our radial-velocity, photometric and Ca II H+K
        index time series are available in electronic format from 
        the CDS through anonymous ftp to cdsarc.u-strasbg.fr 
        or through http://cdsweb.u-strabg.fr/cgi-bin/qcat?J/A+A/
}}

        \subtitle{XVIII. An Earth-mass planet in the GJ 581 planetary system}

\author{M.~Mayor     \inst{1}
         \and X.~Bonfils   \inst{2,3}
         \and T.~Forveille  \inst{2}
         \and X.~Delfosse \inst{2}
         \and S.~Udry         \inst{1}
         \and J.-L.~Bertaux \inst{4}
         \and H. Beust \inst{2} 
         \and F.~Bouchy    \inst{5}
         %\and M.~Gillon \inst{1,8}
         \and C.~Lovis       \inst{1}
         \and F.~Pepe        \inst{1}
         \and C.~Perrier     \inst{2}
         \and D.~Queloz    \inst{1}
         \and N.~C.~Santos \inst{1,6}
%   \and....
}

        \offprints{M. Mayor}
        \institute{
        Observatoire de Gen\`eve, Universit\'e de Gen\`eve, 
          51 ch. des Maillettes,
          CH-1290 Sauverny, Switzerland
          \email{michel.mayor@obs.unige.ch}
        \and
        Laboratoire d'Astrophysique, Observatoire de Grenoble, 
          Universit\'e J.Fourier, CNRS (UMR5571), 
          BP 53, F-38041 Grenoble, Cedex 9, France
        \and
        Centro de Astronomia e Astrof{\'\i}sica da Universidade de Lisboa,
          Observat\'orio Astron\'omico de Lisboa, Tapada da Ajuda, 1349-018
          Lisboa, Portugal
        \and
        Service d'A\'eronomie du CNRS, BP 3, 91371
	  Verri\`eres-le-Buisson, France
        \and
        Institut d'Astrophysique de Paris, CNRS, 
	  Universit\'e Pierre et Marie Curie,
	  98bis Bd Arago, 75014 Paris, France
        \and
        Centro de Astrof{\'\i}sica, Universidade do Porto, Rua das Estrelas,
	  P4150-762 Porto, Portugal
        }

        \date{Received / Accepted }

        \abstract {The GJ~581 planetary system was already known to 
harbour three planets, including two super-Earths planets which 
straddle its habitable zone. We report here the detection of an 
additional planet -- GJ 581e -- with a minimum mass of 1.9 $M_\oplus$. 
With a period of 3.15 days, it is the innermost planet of the system 
and has a $\sim$5\% transit probability.

We also correct our previous confusion of the orbital period of GJ~581d 
(the outermost planet) with a one-year alias, thanks to an extended
time span and many more measurements. The revised period is 66.8 days,
and locates the semi-major axis inside the habitable zone of the 
low mass star.

The dynamical stability of the 4-planet system imposes an upper bound
on the orbital plane inclination. The planets cannot 
be more massive than approximately 1.6 times their minimum mass.}


\titlerunning{An Earth-type planet in \object{GJ~581} planetary system}

\authorrunning{M. Mayor et al.}

\maketitle

%________________________________________________________________

\section{Introduction}
HARPS is a vacuum spectrograph designed to measure precise radial 
velocities, with the specific goal of searching for exoplanets in 
the Southern hemisphere \cite{Mayor2003}. This high-resolution 
Echelle spectrograph (R=115000) is fiber-fed by the ESO 3.6-meter 
telescope at La Silla Observatory. The consortium which built this 
instrument was granted 500 observing nights over five years to 
conduct a comprehensive search for exoplanets. This large programme 
addresses several key exoplanet questions, including the characterization
of very low mass exoplanets. A significant fraction of the observing 
time was devoted to the study of planets orbiting stars at the
bottom of the main sequence. The M-dwarf sub-programme includes over
100 stars, which form a volume-limited sample.

Our HARPS search for southern exoplanets orbiting M-dwarfs is on-going,
and has been recently expanded to a larger sample of some 300 low-mass 
stars. Several planets have already been detected in the M-dwarf 
survey: \object{GJ 581 b} \citep{Bonfils2005b}, \object{GJ 581 c} 
and d \citep{Udry2007}, \object{GJ 674} \citep{Bonfils2007}, 
\object{GJ 176} \citep{Forveille2009}, and here a fourth planet in the
\object{GJ 581} system. These 6 planets all have minimum masses 
under approximately 15 Earth-masses. They represent approximately half 
of the known inventory of planets orbiting M stars, and most of its
lowest mass members. Besides the HARPS planets, this inventory 
includes three planets around \object{GJ 876} \citep{Delfosse1998, 
Marcy1998, Marcy2001, Rivera2005}, a single Neptune-mass planet 
orbiting \object{GJ 436} \citep{Butler2004}, jovian planets bound 
to \object{GJ 849} \citep{Butler2006} and to \object{GJ 317} 
\citep{Johnson2007}, and finally a gaseous giant planet orbiting 
\object{GJ 832} \citep{Bailey2009}. Altogether, 8 planetary systems 
centered on M dwarfs have been identified  by the radial velocity 
technique, for a total of 12 planets. Although the statistics 
are still limited, and come from surveys with different sensitivities, 
they indicate that multi-planet systems are common: the fraction 
of known multiple systems around M dwarfs is already 2/8 (25\,\%),
and it reaches 5/8 (63\,\%) if one accounts for systems where a
radial velocity drift indicates an additional long-period planet.

The competing teams which monitor M dwarfs with Doppler spectroscopy 
have together observed over 300 stars for several years. Planetary 
systems with at least one identified gaseous giant planet
have been found around $\sim$1.5\% of these low mass stars,
and that proportion is clearly lower than for K and G dwarfs. 
Up to periods of several hundred days this comparative
deficit of Jupiter-mass planets for M dwarfs is statistically robust
\citep{Bonfils2006, Endl2006, Johnson2007}. Conversely, planets less massive
than $\sim$25 $M_\oplus$ are significantly more frequent around
M dwarfs \citep[][and 2009 in preparation]{Bonfils2007}.

The GJ~581 system is of particular interest, as two of its previously
detected planets are located on the two edges of the habitable zone (HZ)
of the M3V host star \citep{Selsis2007, vonBloh2007}.  Despite the
uncertainties on the exact location of the ``liquid water'' zone for
planets more massive than the Earth, these detections demonstrate that
super-Earths in the HZ can be detected by Doppler spectroscopy, for 
planets orbiting M dwarfs.

In the present paper, we report the detection of a fourth planet in the 
\object{GJ 581} system, 'e', with a minimum mass of 1.94 $M_\oplus$. We 
also correct the period of the outer planet \object{GJ 581d}, and
find that the revised period locates \object{GJ 581d} in the 
habitable zone. In Section~\ref{sect:prop} we briefly recall 
the main characteristics of the host star \object{GJ 581}, and 
of its first three planets. An enlarged set of measurements 
(Section~\ref{sect:data}) allows a reexamination of the structure 
of the GJ581 planetary system, and the discovery of an additional 
very-low-mass planet (Section~\ref{sect:fourth}).
Since Doppler spectroscopy, and all the more so for M dwarfs, can
be confused by stellar surface inhomogeneities, we dedicate 
Section \ref{sect:activity} to discussing the magnetic activity of
\object{GJ 581}. In Section~\ref{sect:dynamic}, we use dynamical 
stability considerations to obtain upper limits to the 
planetary masses. To conclude, we briefly discuss how this system 
meshes up with theoretical predictions, and consider the prospects 
of finding even lower mass planets in the habitable zone of M dwarfs. 

\section{\label{sect:prop}Stellar caracteristics and planetary system}
Our report of the first Neptune-mass planet on a 5.36-d orbit
around GJ 581 \citep{Bonfils2005b} extensively describes the
properties of the host star. Here we will therefore only summarize
those characteristics.

GJ~581 (HIP~74995, LHS~394) is an M3 dwarf at a distance to the Sun of
6.3 pc.  Its estimated luminosity is only 0.013 $L_\odot$. 
\citet{Bonfils2005} find GJ 581 slightly metal-poor ([Fe/H]
= -0.25), in contrast to the supersolar metallicities 
\citep{Santos2001,  Santos2004a} of most solar-type stars 
hosting giant planets. All indirect tracers (kinematics, stellar 
rotation, X-ray luminosity, and chromospheric activity) suggest 
a minimum age of 2~Gyr. The HARPS spectra show weak \ion{Ca}{ii} H and K
emission, in the lower quartile of stars with similar spectral
types. This very weak chromospheric emission points towards a 
long stellar rotation period.

The first planet was easily detected after only 20 HARPS
observations \citep{Bonfils2005b}, and the periodogram of the 
residuals of that first solution hinted at power excess around 
13~days. As 30 additional observations were accumulated for a 
total time span of 1050~days, that coherent signal strengthened 
to strong significance, and a third planet appeared, though still 
with a significant False Alarm Probability (FAP): \object{GJ 581c} 
had $m\sin i = 5 M_\oplus$, $P=12.9$~days) and a $\sim0.28\%$ FAP, 
while the more distant \object{GJ 581d} had 
$m \sin i= 7.7 M_\oplus$, $P = 83$ days, and a $\sim3\%$ FAP.
\citep{Udry2007}. \citet{Udry2007} therefore ended on a call for 
confirmation by additional radial velocity measurements, and 
based on the very-low magnetic activity of \object{GJ 581}, 
suggested that the significant residuals might reflect
additional planets in the system.

\section{\label{sect:data}A new data set}

Following up on that call for more data, we have now recorded a
total of 119 \textsc{Harps} measurements of \object{GJ 581} 
(Table~\ref{tab:rv}, only available electronically), 
which together span 1570 days (4.3 years). In parallel to this
extension of the data set, the precision of all measurements, 
including the previously published ones, has been improved. 
The reference wavelengths of the thorium and argon lines of
the calibration lamps were revised by \citet{Lovis2007}, and 
most recently Lovis (2009, in prep.) implemented a correction 
for the small effect of Th-Ar calibration lamp aging. The 
changing internal pressure of aging lamps shifts the wavelength 
of their atomic transition. The effect is now corrected down to 
a level of just a few 0.1 m/s, using the differential pressure
sensitivity of the argon and thorium lines. The HARPS pipeline 
also corrects the radial velocities to the Solar System Barycentric 
reference frame \citep{Lindegren2003}. That correction is computed
at the weighted-mean time of the exposure, which is measured by a
photometer that diverts a small fraction of the stellar flux. 
This effective time of the exposure is measured to better than 
1\%, which at the latitude of La Silla Observatory and for 
900s-long exposures translates into a $\sim0.3~\mathrm{m\,s^{-1}}$
worst case error on the Earth-motion correction. The combined
photon noise, calibration error, and error on the mean time of
the exposure result in a typical overall uncertainty of 
$1~\mathrm{\mathrm{m\,s^{-1}}}$, with a full range of
0.7 to 2~$\mathrm{m\,s^{-1}}$ which reflects the weather
conditions. Finally, the pipeline corrects for the 
$0.20~\mathrm{\mathrm{m\,s^{-1}\,yr^{-1}}}$ perspective acceleration 
from the proper motion of the star \citep[e.g.][]{Kurster2003}.

\addtocounter{table}{1}

\section{\label{sect:fourth}GJ 581, a system with four low-mass planets}

Our preferred method for planet searches in RV data uses a heuristic 
algorithm, which mixes standard non-linear minimizations with genetic 
algorithms. It efficiently explores the large parameter space of 
multi-planet systems, and quickly converges toward the best solution. We 
tried models composed of zero to five planets on Keplerian orbits, and 
found that the data require a 4-planets model.  

To verify the robustness of that 'black-box' solution, and gain insight
on its content, we also 
performed a step by step periodogram analysis. The top panel of 
Fig.~\ref{fig:periodo} displays the window function of our RV 
measurements, and unsurprisingly shows that the dominant periodicities 
of our sampling are 1~day and 1~year. The second panel is a floating-mean 
periodogram \citep{Gilliland1987, Zechmeister2009} of the RVs. It is
dominated by a $\sim$5.36-day periodicity, which corresponds to 
\object{GJ 581b}. To estimate the FAP of that peak, we generated 
virtual data sets using boostrap resampling \citep{Press1992} of the
actual measurements, and examined the peak power in the periodogram
of each of these virtual datasets. None of 10,000 trials had as high
a peak as measured for the actual 5.36-day signal, which therefore 
has a FAP below 0.01\%.  We then adjusted a Keplerian orbit with that
starting period, and examined the periodogram of the residuals of
this 1-planet solution (Fig.~\ref{fig:periodo}, 3$^{rd}$ panel). That
second periodogran is dominated by a peak around $\sim12.9$ day, with 
a FAP$<$0.01\%, which corresponds to the known planet \object{GJ 581c}. 
We then investigated the periodogram of the residuals of the 
corresponding 2-planets least square adjustment. That new periodogram
has power excess in the 50-90 days range and around 3.15 days (as well
as at 1-day aliases of these periods), both with FAPs$<$0.01\%. 
The broad power excess between 50 and 90 days splits into 3 separate 
peaks at $\sim$59, 67 and 82 days, which are 1-year aliases of each
other ($1/67-1/365\sim1/82, 1/67+1/365\sim1/59,)$. In \citet{Udry2007}, 
we attributed an 82-days period to \object{GJ 581 d}. The corresponding 
periodogram peak now has markedly less power than the 67-days periodicity,
and attempts to adjust orbits with that period (as well as with a 59~days
period) produced significantly larger residuals. We conclude that the
actual period of \object{GJ 581 d} is around 67~days and that we had
previously been confused by a 1-year alias of that period. As we discuss
below, the revised period locates \object{GJ 581 d} in the habitable zone
of \object{GJ 581}. Subtracting 
the corresponding 3-planet adjustement leaves no power excess in the 
50-90 days range (5$^{th}$ panel), and markedly enhances the 3.15 day peak.
The residuals of this 4-planet solution finally display no statistically 
significant peak (6$^{th}$ panel). The strongest peak is located close to 
one year ($P\sim384$ day) and has a $\sim$50\% FAP.

Both the heuristic algorithm and the spectral analysis method
find the same 4 coherent signals in our data.  Once this best
approximate 4-planet solution had been identified, we performed a final
global least square adjustment, using the iterative Levenberg-Marquardt 
algorithm. Since we find insignificant eccentricities for planets 'b' 
and 'c', we fixed the eccentricities and longitude of  
periastron to zero for both planets. Table~\ref{TabOrb2} lists the 
resulting orbital elements, and Fig.~\ref{fig:orbits} displays 
the corresponding Keplerian orbits.

Compared to our \citet{Udry2007} study, we thus add a fourth planet and
very significantly revise the characteristics of the outer planet 
(\object{GJ 581d}), and in particular its period. Planets $b$ to 
$e$ have respective minimum masses of $m \sin i \sim$~15.7, 5.4, 
7.1 and 1.9 $M_\oplus$. \object{GJ 581e} is the lowest
mass exoplanet detected so far around a main sequence star. 
None of the period ratios of the system comes close to any simple 
ratio of integers, with e.g. P$_b$/P$_e$ = 1.70, P$_c$/P$_b$ = 2.4 and
P$_d$/P$_c$ = 5.17. Non-resonant hierarchies have similarly been
noticed for all multi-planet systems which only contain low mass planets,
such as HD 40307 \citep{Mayor2009} and HD 69830 \citep{Lovis2006}.

\begin{table*}[t!]
\caption{\label{TabOrb2}Fitted orbital solution for the GJ\,581 
  planetary system: 4 Keplerians. The model imposes circular orbits
  for planets \object{GJ 581 b} \& e, since the derived eccentricities 
  for a full Keplerian solution are insignificant (see text).}

\begin{tabular}{l l l c c c c}
  \hline\hline
  \multicolumn{2}{l}{\bf Parameter} &\hspace*{2mm} 
  & \bf GJ\,581\,e  & \bf GJ\,581\,b  & \bf GJ\,581\,c & \bf GJ\,581\,d \\
  \hline
  $P$ & [days] & & 3.14942 $\pm$ 0.00045 & 5.36874 $\pm$ 0.00019 & 12.9292 $\pm$ 0.0047 & 66.80 $\pm$ 0.14\\
  $T$ & [JD-2400000] & & 54716.80 $\pm$ 0.01 &  54712.62 $\pm$ 0.02 & 54699.42 $\pm$ 0.87 & 54603.0 $\pm$ 2.2 \\
  $e$ & & & 0 (fixed) & 0 (fixed) & 0.17 $\pm$   0.07 & 0.38 $\pm$   0.09  \\
  $\omega$ & [deg] & & 0 (fixed)  & 0 (fixed) & -110 $\pm$ 25 & -33 $\pm$  15\\
  $K$ & [m s$^{-1}$] & & 1.85 $\pm$ 0.23 & 12.48 $\pm$ 0.23 & 3.24 $\pm$ 0.24 & 2.63 $\pm$ 0.32\\
  $V$ & [km s$^{-1}$] & & \multicolumn{4}{c}{-9.2082 $\pm$ 0.0002}  \\
  $f(m)$ & [10$^{-14} M_{\odot}$] & & 0.21 & 108.11 & 4.34 & 10.05\\
  $m_2 \sin{i}$ & [$M_{\oplus}$] & & 1.94 & 15.65 & 5.36 & 7.09 \\
  $a$ & [AU] & & 0.03 & 0.04 & 0.07 & 0.22 \\
  \hline
  $N_{\mathrm{meas}}$ & & & \multicolumn{4}{c}{119} \\
  {\it Span} & [days] & & \multicolumn{4}{c}{1570} \\
  $\sigma$ (O-C) & [ms$^{-1}$] & & \multicolumn{4}{c}{1.53} \\
  $\chi^2_{\rm red}$ & & & \multicolumn{4}{c}{1.49} \\
  \hline
\end{tabular}

\end{table*}

\begin{figure}
\centering
\includegraphics[width=0.9\linewidth]{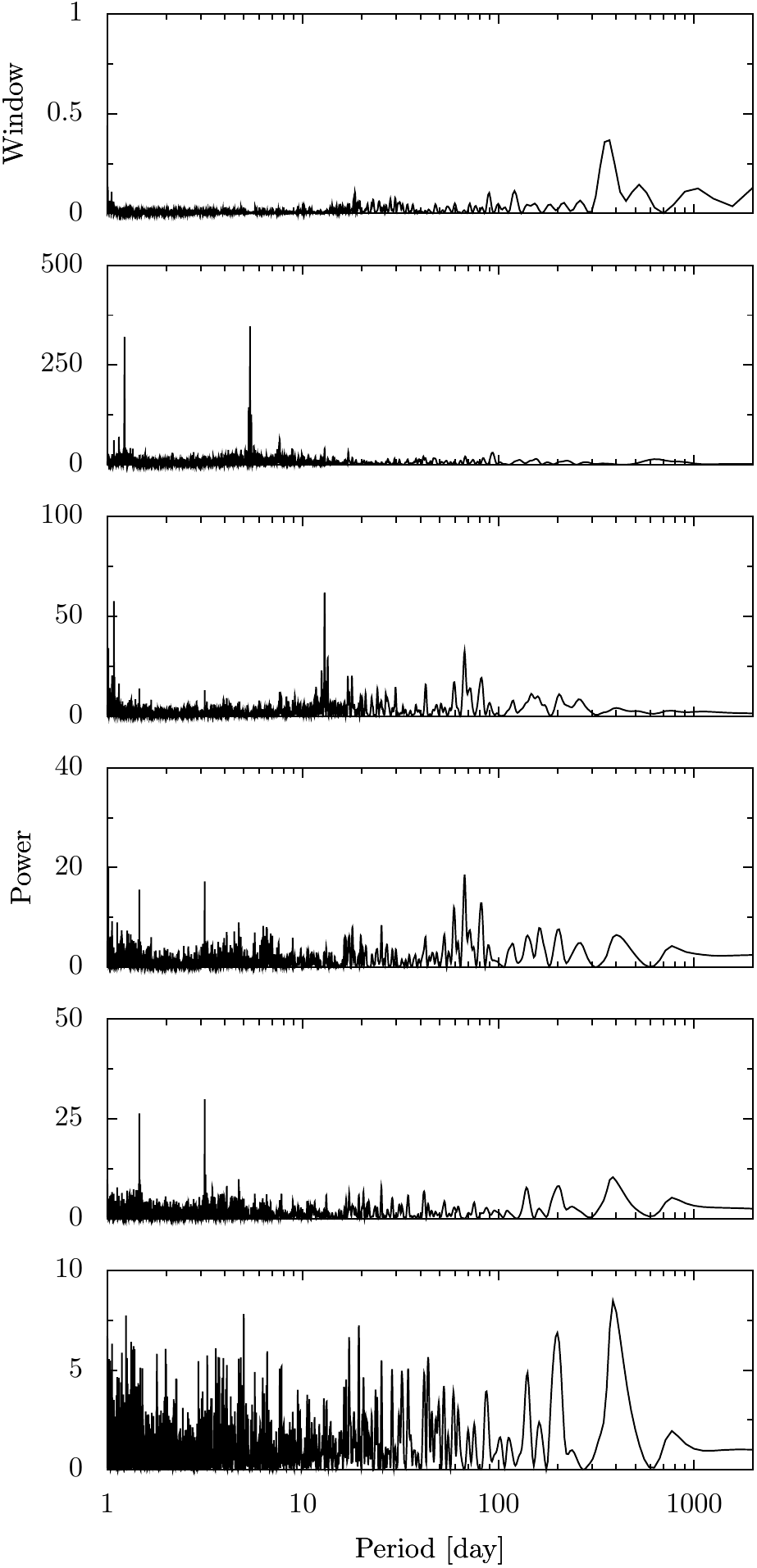}
\caption{\label{fig:periodo} Periodograms of GJ581 radial velocity
  measurements. {\it Panel 1 (top)}: Window function of the HARPS
  measurements. {\it Panel 2}: Periodogram of the HARPS
  velocities. The peaks corresponding to GJ581b (P=5.36d) and its alias
  with one sidereal day are clearly visible. {\it Panel 3}: HARPS
  velocities corrected from the velocity variation due to GJ581b. The
  peak corresponding to GJ581c (P=12.9d) appears. {\it Panel 4}:
  Removing the variability due to planets b and c, we can see the
  peak due to the P=3.15d variability of GJ581e (+ the alias with the
  a one sideral day period) as well as the peak at 66d due to GJ581d. 
  {\it Panel  5} : Periodogram after removing the effect of planets b, c and
  d. The peak due to planet e (Mass = 1.9 earth-mass) is evident (+
  its one-day aliases). {\it Panel 6 (bottom)}: Removing
  the velocity wobbles due to the four planets, the periodogram of
  the residual velocities shows no additional significant
  peaks.  }
             \label{fig:periodo}
\end{figure}
\begin{figure}
\centering
\includegraphics[width=0.9\linewidth]{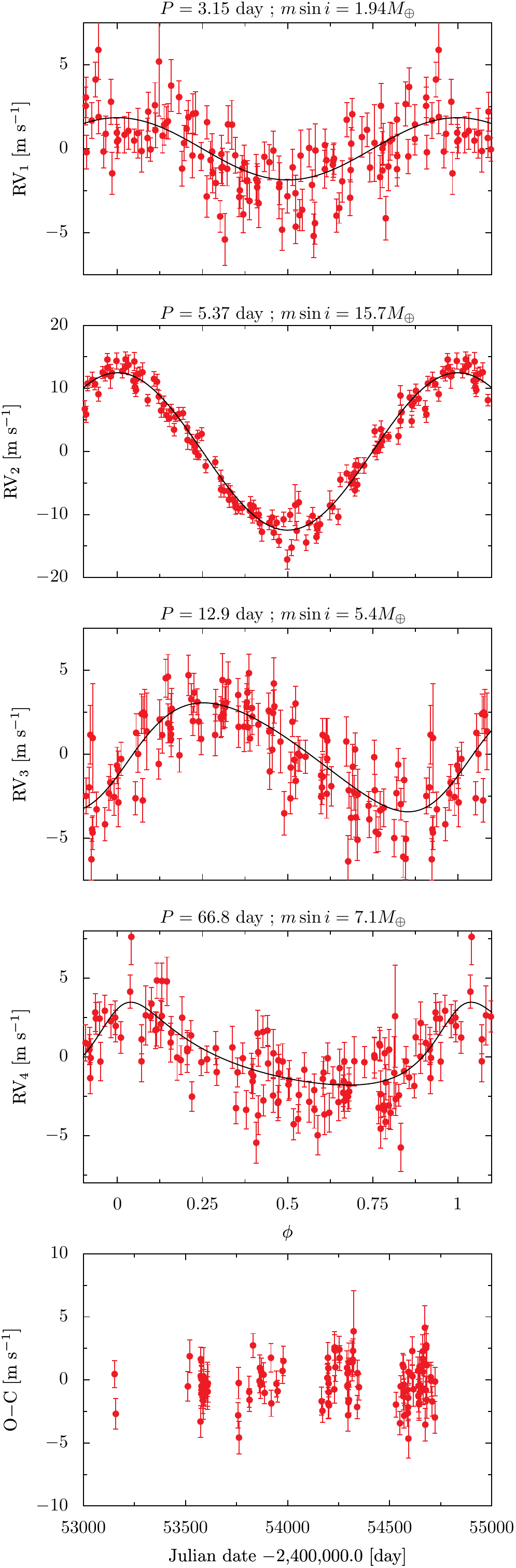}
\caption{\label{fig:orbits} Radial velocity curves for planets e, b, c
  and d, from top to bottom. The lowest panel displays the residual 
  to the four-planets keplerian fit. }
             \label{fig:orbits}
\end{figure}

\section{\label{sect:activity}Planets rather than spots}

Coherent Doppler shifts unfortunately do not always correspond to 
planets. Inhomogeneities of the stellar surface like spots, plages, flares or
convective patterns can break the even distribution between the blue- and
red-shifted halves of a rotating star and induce apparent radial velocity
shifts. At best, this introduces noise on the RV measurements, which is 
usually referred to as ``jitter''. In some cases the surface pattern 
can be stable over long time spans (from days to years), and the
``jitter'' noise then is correlated. In such cases, it does not average
out as white noise would, and instead builds up a coherent signal.
That apparent Doppler shift can easily happen to mimic
a Keplerian orbit \citep[e.g.][]{Queloz2001, Bonfils2007}.

M dwarfs notoriously tend to have stronger magnetic activity than 
solar-like stars, and possible false detections from coherent jitter 
therefore need even closer scrutiny. We have indeed encountered several
such cases in our M star survey, including \object{GJ 674} 
\citep{Bonfils2007} and \object{GJ 176} \citep{Forveille2009}.
In both cases, photometric monitoring and spectroscopic activity
diagnostics (measuring the \ion{Ca}{ii} H\&K and H$_\alpha$ lines) 
demonstrated that a seemingly Keplerian signal was in fact due 
to spot modulation of the absorption line profiles. These two examples 
serve as reminders that a periodic Doppler signal does not always 
reflect the presence of a planet, and that additional checks are 
needed to conclude.

In our previous papers on the \object{GJ 581} system, we could 
exclude any such confusion for the first three planets. The 
$\sim13 \mathrm{m\,s^{-1}}$ radial velocity amplitude due to
\object{GJ 581 b} \citep[$P=5.36$ day --][]{Bonfils2005b} is
sufficiently large that a spot signal of that magnitude would 
have induced measurable distortions of the rotational profile.
The lack of any correlation between the bisector span (a first 
order measurement of spectral lines asymmetry) and the measured RVs,
excludes that scenario. The strength of the \ion{Ca}{ii} 
H\&K emission lines in addition classifies \object{GJ 581} in the
lowest $\sim5$\% magnetic activity bin of our M dwarfs 
sample. This implies a much longer rotation period than the observed
5.4~days, as confirmed by the very low projected rotational velocity 
$v \sin i < 1~\mathrm{km\,s^{-1}}$). That velocity would need 
an unprobably low inclination $i$ to be compatible with a 5~days
rotation period, The additional measurements since 
\citet{Bonfils2005b} have strengthened those conclusions even
further.

In \citet{Udry2007}, we used distinct lines of arguments for the 12.9
and the 83 days (now 67 days) signals. The arguments developed 
above for \object{GJ 581b} hold for the 12.9-day period, which is
also too short to be compatible with the rotation period. They
also apply for \object{GJ 581e} at 3.1 days. An 83 days
(or 67 days) period, on the other hand, would be perfectly consistent 
with the rotation period of an M dwarf with the activity level 
of \object{GJ 581}. For such long periods the bisector span also 
loses its  diagnostic power, because the rotational profile becomes
much too narrow to resolve with HARPS. This calls for a different
argument, and we used a relation by \citet{Saar1997} to estimate
the minimum filling factor of a spot that could produce the observed
radial velocity signal. That relation,
\begin{equation}
K_s \sim 6.5 \times f_s^{0.9} \times v \sin i \ \ \ \ [\mathrm{m\,s^{-1}}]
\end{equation}
connects the semi-amplitude $K_s$ of the radial-velocity signal 
induced by an equatorial spot with its surface filling factor, 
$f_s$\%, and the projected rotation velocity, $v \sin i$. 
It has been validated in numerous examples, including
e.g. \object{GJ 674} and \object{GJ 176}. 
Applying that relation to a $R_\star=0.31~\mathrm{R_\odot}$ stellar 
radius, a $P=83$ days period and a $K\sim2.5~\mathrm{m\,s^{-1}}$ 
semi-amplitude, Eq. (1) readily showed that the \object{GJ 581 d} signal
would have needed a $f_s\sim2.2\%$ filling factor to be explained
by a spot. Such a large spot is firmly excluded by the sub-\% photometric
stability of \object{GJ 581}.

Our revised 66.7 days period for \object{GJ~581 d} slightly weakens that 
argument, since a shorter period signal can be induced 
by a smaller spot. The revision however is minor, as the 67-day
signal still requires a $f_s=1.7\%$ spot, which is easily excluded 
by the improved photometric measurements that have become 
available since \citet{Udry2007}. The strongest photometric 
stability constraint is provided by 6 weeks of continuous 
monitoring of GJ~581 by the MOST satellite. Preliminary 
results, presented by J.~Matthews at the 2007 Michelson Summer
School\footnote{http://nexsci.caltech.edu/workshop/2007/Matthews.pdf},
indicate that the peak-to-peak photometric variability during those 
6~weeks was less than 5~mmag. The 66.7-day HARPS radial velocity signal
is therefore incompatible with rotational modulation of spots or 
granulation patches.

The variations of the chromospheric \ion{Ca}{ii} H$+$K line 
(Table~\ref{tab:rv}, only available electronically) provide
another diagnostic of magnetic activity, which proved very valuable
for \object{GJ 674} \citep{Bonfils2007}. We parametrize this 
chromospheric emission by
\begin{equation}
Index = \frac{H+K}{R+B},
\end{equation}
where H (resp. K) is the flux measured in the \ion{Ca}{ii} H (resp. K)
line and where R and B measure the flux in pseudo-continuum 
bands on both sides of the lines.
Fig.~\ref{fig:caii} displays our \ion{Ca}{ii} H$+$K measurements as 
a function of time (top panel), and the corresponding periodogram
(lower panel). The figure suggests long term variations, over
timescales of several thousands days which are much too long 
to be the rotational period. If those variation are real, they 
would be more likely to reflect an analog of 11~years solar 
magnetic cycle. Magnetic cycles can potentially affect apparent 
stellar velocities, since they can change the balance between
ascending (blue-shifted) and receding (red-shifted) convective
elements in the atmosphere, but we see no evidence for long-term 
RV variations. The other significant structure in the 
chromospheric signal is a broad power excess between 80 and 
120 days, which is much more likely to reflect the rotation 
period. That peak overlapped our incorrect initial estimate 
of the period of \object{GJ 581 d}, but as discussed above 
photometry limits the amplitude of any velocity signal from 
spots at that period to at most a quarter of the observed value.
The revised period puts a final nail in the coffin of that concern,
since the periodogram of the \ion{Ca}{ii} H$+$K indices has no power
excess around 67 days. The Ca~II H~+~K periodogram finally has a 
weak peak at $\sim15$~days, which does not match any of the observed 
radial-velocity periods.

\begin{figure}
\centering
\includegraphics[width=0.9\linewidth]{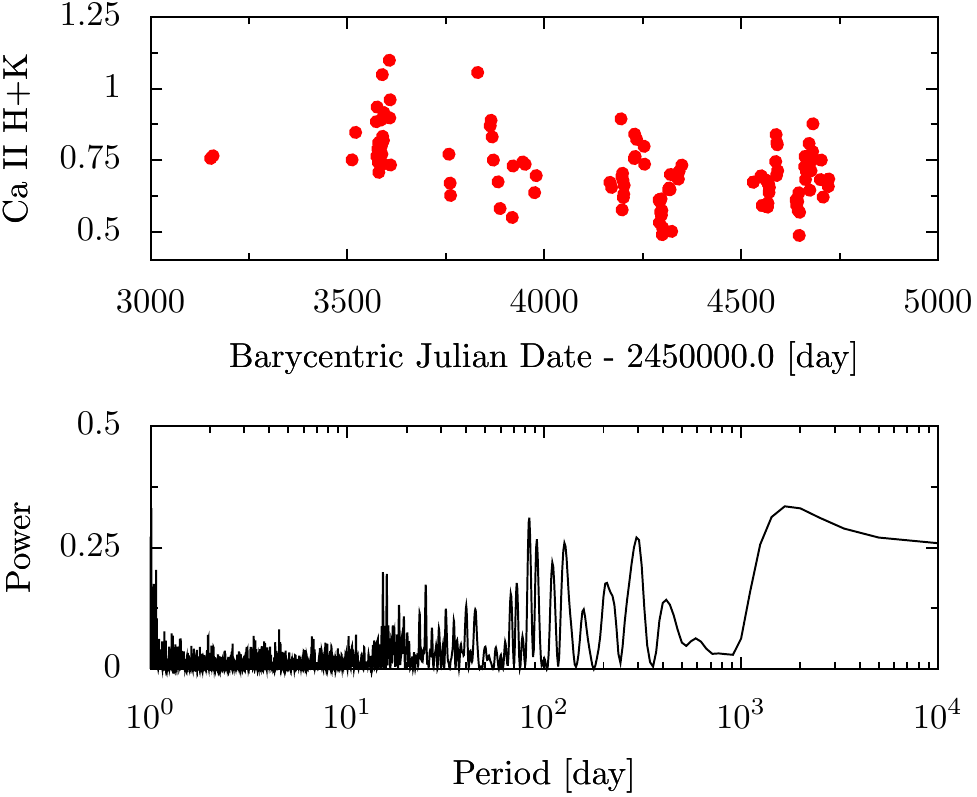}
\caption{\label{fig:caii}The \ion{Ca}{ii} H$+$K index as function of
  Julian date (upper panel) and its periodogram (lower panel).
}
             \label{fig:caii}
\end{figure}

To summarize, none of the 4 periodic signal detected in our RV
measurements is compatible with rotational modulation of stellar
surface patterns. Four low-mass exoplanets orbiting the star 
remain, by far, the most likely interpretation.

\section{\label{sect:dynamic}Dynamical evolution}
We reinvestigate the dynamics of the revised \object{GJ 581} system,
using the same techniques as in \citet[][B08]{Beust2008}. 
With the SyMBA symplectic $N$-body code \citep{Duncan1998}, 
we carry 5-body integrations over 0.1 Gyr, with a $2\times10^{-4}\,$yr
timestep. Starting from the nominal solutions (Tab.\ref{TabOrb2}), 
which assumes initially circular orbits for Gl581e and Gl581b (the 
two innermost planets), we performe those integrations for 
inclinations of 90 to 10 degrees, assuming full coplanarity of 
all orbits.

Our nominal simulation uses the periastron parameters derived in the
Keplerian fit and listed in Table~1. 
To probe the sensitivity of 
the simulations to these parameters,  we performed additional
integrations with different starting values for the periastron 
longitudes. Unsurprisingly, since the orbits are non-resonant,
the results were not significantly different.

\begin{figure*}
\makebox[\textwidth]{
\includegraphics[width=0.49\textwidth]{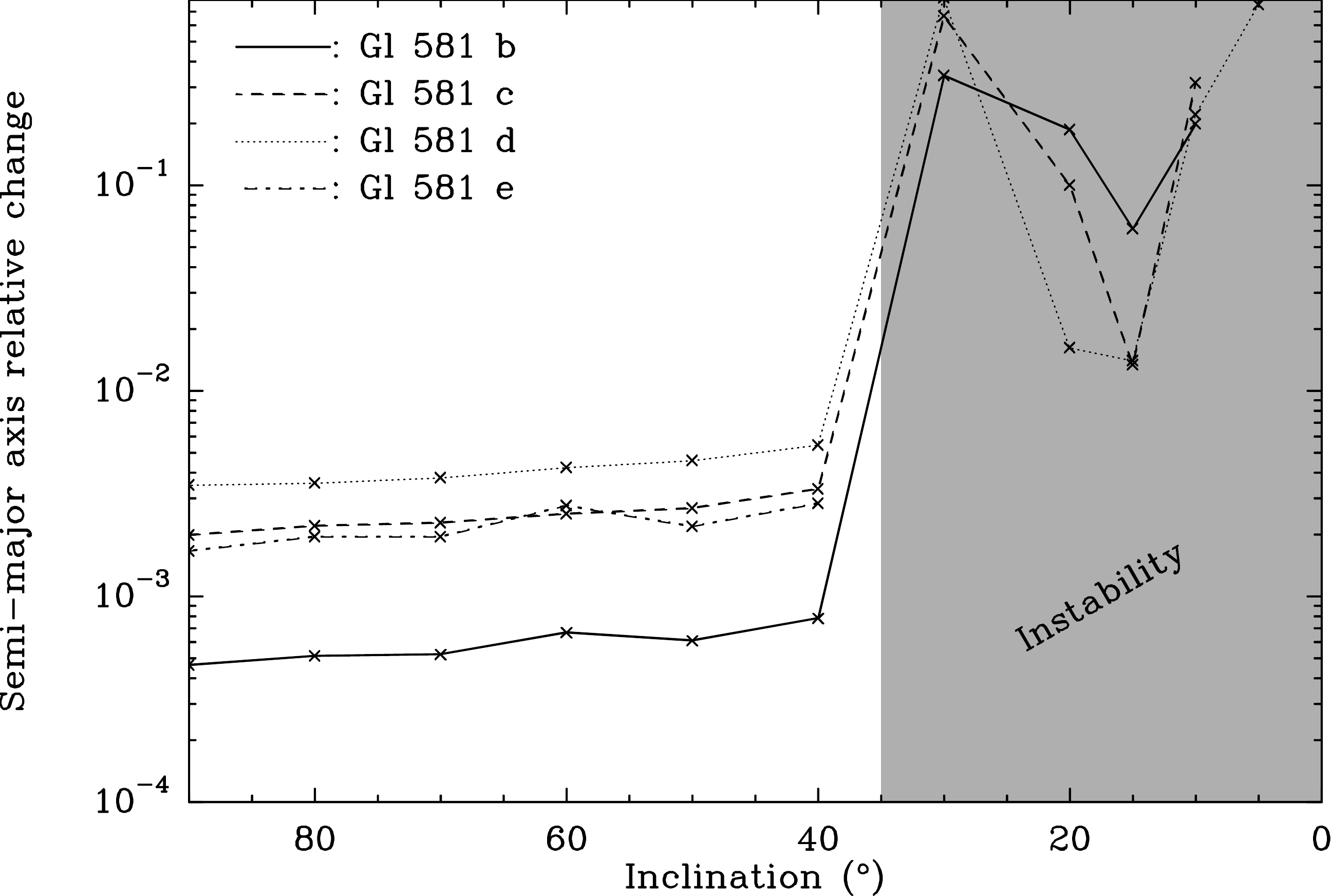} \hfil
\includegraphics[width=0.49\textwidth]{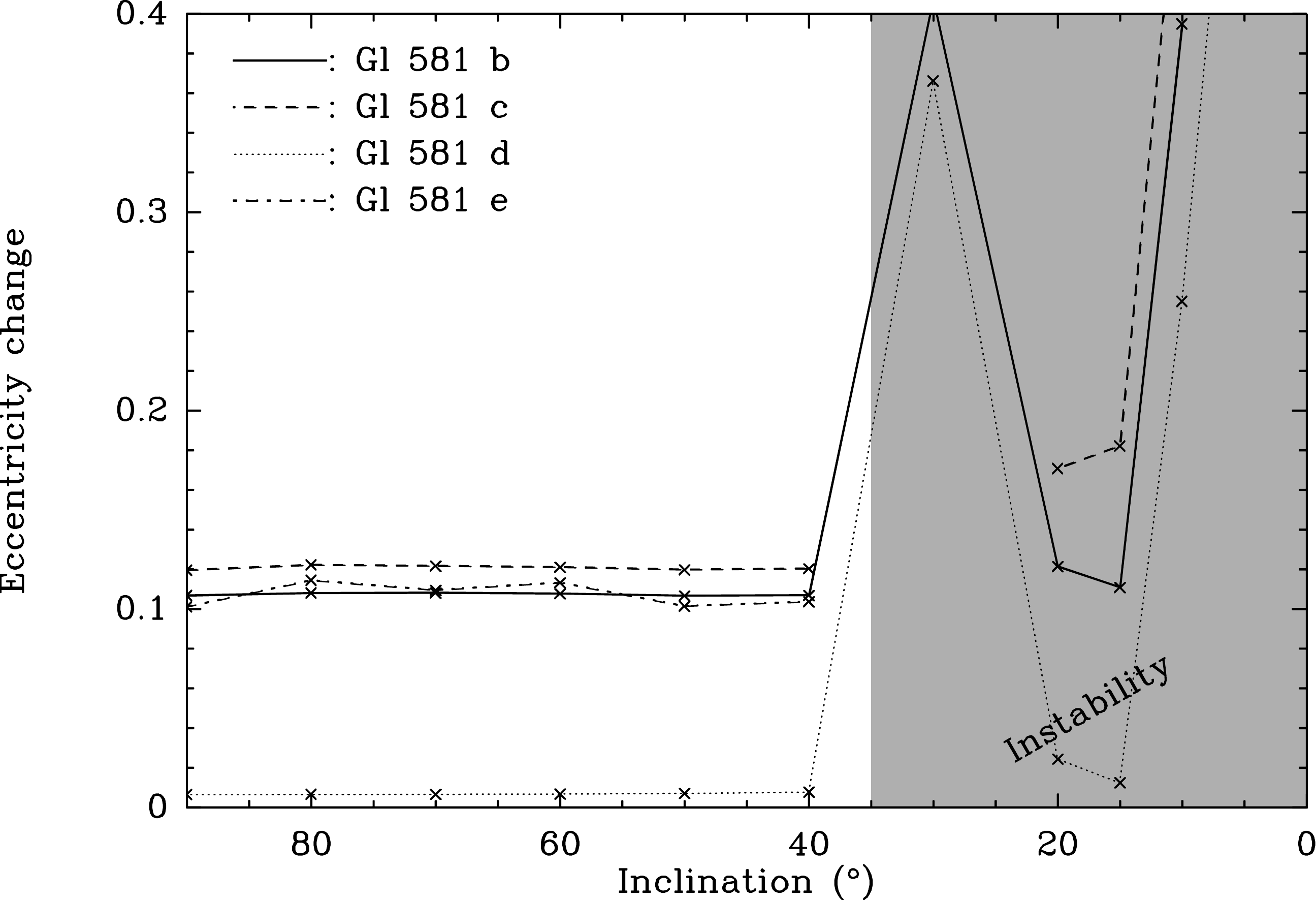}}
\caption[]{Stability of various configurations of the four-planet 
  Gl~581 system. The left and right panels respectively show the 
  maximum variations, after $\times10^8$ years, of the semi-major axes
  and of the eccentricities (right plot) as a function of the 
  assumed inclination of the system (i=0 is pole-on). Each cross 
  represents one simulation. The grey shading for $i<35\degr$ 
  marks an instable zone where one or more planets were ejected 
  during the integration.
 }
\label{incli}
\end{figure*}

Figure~\ref{incli} shows the maximum variations of the semi-major axes 
and eccentricities during those integrations. As in B08, we note that 
the semi-major variations gradually increase as the inclination decreases. 
This is a clear indication that lower inclinations, i.e. higher 
planetary masses, produce a less stable system. Compared to our 
previous study, the addition of \object{GJ 581e} drastically tightens 
the minimum inclination constraint. That low mass planet has little
effect on the stability of the more massive \object{GJ 581b}, 
\object{GJ 581c}, and \object{GJ 581d}, but for inclinations 
$5\le i\le30\degr$, \object{GJ 581e} is ejected out of the system 
on time scales of at most a few Myrs.

The nominal integrations assume circular initial orbits for 
\object{GJ 581e} and \object{GJ581b}. Relaxing this constraint 
and setting the initial eccentricity of planet e to 0.1, we 
found the system even less stable: \object{GJ 581e} was then ejected 
for inclinations up to $i=40\degr$.

Unlike most multi-planet radial-velocity systems, the nominal least
square adjustment to the radial velocities is dynamically stable
for sufficiently high inclinations (and co-planar orbits). Restricting 
the solutions to dynamically stable systems only imposes that 
$i\ga40\degr$ and therefore that the mass of each planet be
no more than $\sim$1.6 times its minimum mass. For \object{GJ 581e}, 
b, c and d, those upper limits are 3.1, 30.4, 10.4 
and 13.8 $M_\oplus$.

\section{Summary and discussion}

The large number of precise HARPS measurements has now revealed 
four planets in a non resonnant-configuration. The false alarm
probablities of all four planetary signals are below 0.01\%. The
orbital elements of planets GJ581b \citep[P=5.36d][]{Bonfils2005} 
and GJ581c \citet[P=12.9d][]{Udry2007} are fully confirmed, and
slightly refined. The 82 days \citet{Udry2007} period for \object{GJ 581d},
on the other hand, was a one-year alias of the true 66.8 days
period of that planet. Our extended span and much larger number 
of measurements correct that confusion. 
Photometry, as well as monitoring of the \ion{Ca}{ii} H$+$K activity 
indicator over the four years of our spectroscopic measurements,
indicate that the signal at 66 days does not originate in stellar 
activity. At the present time, and based on the well defined radial 
velocity signal and on an estimate of the worst-case effect of 
stellar activity, a planet with a period of 66.8 days is by far 
the most probable interpretation.

The new orbital elements for the outer planet GJ 581d correspond to 
a 0.22~AU semi-major axis. A planet with an eccentric orbit receives 
a larger orbit-averaged flux than a planet on a circular orbit of 
the same semi-major axis \citep{Williams2002}, and GJ~581 d receives 
the same average flux as a planet on a a~=~0.21~AU circular orbit.
The \citet{Udry2007} orbital elements of \object{GJ 581d},
with a 0.25~AU semi-major axis, located the planet close to the 
outer edge of the habitable zone, but most likely just slightly outside 
\citep[e.g.][]{Selsis2007}. The revised parameters locate \object{GJ 581d}
firmly in the habitable zone.

Thanks to 119 precise radial velocity measurements, we have discovered 
a fourth planets in the \object{GJ 581} system. Its minimum mass, 
$m \sin i = 1.9 M_\oplus$, is by far the lowest amongst planets 
detected around main sequence stars. Such a low-mass planet is almost 
certainly rocky, and its equilibrium temperature is too high 
to allow a substantial atmosphere.

\object{GJ 581} has the largest number of measurements in our HARPS 
M-star sample, which evidently helped detect 
such a low mass planet in a four planet system. The detection however
does not stretch the limits of Doppler spectroscopy, and for the same
period a planet only half as massive would have been detectable in our 
dataset. Our HARPS measurements set an upper limit of 1.2 m/s on the
stellar jitter of \object{GJ 581}, based on a quadratic subtraction
of just the mean photon noise from the observed dispersion of the 
radial-velocity residuals to the 4-planet solution. Some small 
fraction of this value could still be instrumental, but a 1.2 m/s
``jitter'' allows  the detection of a 2 $M_\oplus$ planet with a
25 days orbital period. Such a planet would be earth-type and in
the middle of the habitable zone of its M dwarf.

The stellar ``jitter'' has different physical components, with distinct 
time scales.  For more massive stars and shorter integration times, 
acoustic modes represent a major contribution, but the short acoustic
timescales in this small star and the 15 minutes integration ensure
that they have been averaged out to below a few 0.1 m/s. The 1.2~m/s
jitter therefore mostly reflects longer time-scale phenomena, such 
as granulation noise and changes in stellar surface anisotropies 
during a magnetic cycle. We are developing optimized measurement 
strategies to average out the granulation noise contribution 
to the stellar jitter, but HARPS already demonstrates that 
the global jitter (including instrumental errors, but not photon noise) 
of reasonably chromospherically quiet M~dwarfs is at most of the
order of 1 m/s. 

This new low mass planet adds to the small set of Doppler-detected 
super-Earths, planets with a minimum mass under 10~$M_\oplus$.  
This planetary mass domain, as well as the Neptune-mass range
($<$30 $M_\oplus$) has largely been populated by the HARPS surveys.
A few statistical properties have already emerged from these early 
discoveries \citep{Mayor2008}:

\begin{itemize}
\item[$-$] The full distribution of planetary masses is bimodal,
with distinct peaks corresponding to gaseous giants and 
super-Earths. Despite the observational bias against low mass planets,
the distribution below $\sim$30 M$_\oplus$ rises towards super-Earth planets 
\citep[cf. Fig.~7 in][]{Mayor2008}.

\item[$-$] The majority of super-Earths and Neptunes are found in 
multiplanetary systems. Four of the 6 planetary systems with a known 
super-Earth, (\object{GJ 876}, \object{HD 40307}, \object{HD 7924}, 
\object{GJ 176}, \object{GJ 581}, \object{HD 181433}) have multiple
planets. Two of these 4 multi-planet systems associate one super-Earth 
with one or two gaseous giant planets (\object{GJ 876}, \object{HD 181433}),
and the other 2 have several low-mass planets on non resonant orbits 
(\object{HD 40307}, \object{GJ 581}). \object{GJ 176} and \object{HD 7924}
have only one detected planet, though they could obviously have more
which haven't been detected yet. \object{GJ 176} is too magnetically 
active to easily detect a lower-mass planet, but the periodogram of 
\object{HD 7924} does indeed show hints of possible additional planets 
\citep{Howard2009}.

\item[$-$] Contrary to gaseous giants, low mass planets seem no 
more frequent around metal-rich host stars \citep{Udry2006a}.

\item[$-$] In a preliminary analysis, we have detected low mass 
close-in planets (P$<$ 100d and $m \sin i < 30 M_\oplus$) around 
30\% of solar-type stars of the HARPS high precision
survey  (Lovis et al. 2009, in prep.).
\end{itemize}

Systems with several low mass close-in planets provide interesting
constraints for models of planetary formation. Three systems are
particularly noteworthy: \object{HD 69830} (3
planets), \object{HD 40307} (3 planets) and \object{GJ 581} (4 planets).

\citet{Terquem2007}, for instance, studied the migration of cores and
terrestrial planets induced by their interaction with the
protoplanetary disk, and suggest that  ``if hot super-Earths
or Neptunes form by mergers of inwardly migrating cores, then such
planets are most likely not isolated.  We would expect to always find
at least one, more likely a few, companions on close and often
near-commensurable orbits''. The high observed fraction of multi-planet
systems matches that prediction, but the observed periods are
quite far from near-commensurability.

\citet{Raymond2008} discuss observable consequences of planet 
formation models in systems with close-in terrestrial planets,
and specifically address the case of \object{GJ~581}. 
In addition to system architecture, they consider information 
from planet composition, which unfortunately is unavailable for 
the \object{GJ 581} system. These authors find two formation 
mechanisms consistent with the available data: {\it in situ} formation,
and formation at a larger distance followed by type I migration. 
{\it In situ} formation however would require a massive disk, 17-50 times 
above the minimum-mass disk. Type I migration from a larger distance,
on the other hand, is expected to leave the close-in low mass planets
in a MMR (mean motion resonance), at odds with the characteristics
of the \object{GJ 581} system.

\citet{Kennedy2008} study hot super-Earths formation as a function 
of stellar mass and suggest that, with migration, short period
low-mass planets are most likely to form around low-mass stars. 
Above approximately 1~solar-mass, the minimum protoplanet 
mass for migration to close-in orbits is above 10 Earth-masses and no
hot super-Earths can form. Our searches for low mass planets with
the HARPS spectrograph will test that prediction.

Zhou et al. 2009 (in prep.) study the birth of multiple super-Earths 
through sequential accretion. Focusing on the formation of the 
\object{HD 40307} triple super-Earths system, they found
evidence that massive embryos can assemble in a gaseous
protostellar disk. The final states of their numerical simulations 
favor models with a reduction factor of ten for type I migration.  These
simulations quite succesfully reproduce the global structure of
the triple super-Earth system, including non-resonant period ratios.

A very recent paper by \citet{Ogihara2009} provides an illuminating
discussion of formation mechanisms of low mass multiple planets orbiting
M stars. It shows, in particular, that the final orbital configuration 
could be quite sensitive to migration timescale. The water fraction 
in low mass planets close to the habitable zone of M dwarfs is 
probably very large, making GJ~581d a serious candidate for 
an ocean planet \citep{Leger2004}.

Our HARPS surveys search for low-mass close-in planets around stars with
masses from 0.3 to over 1~solar-mass. The numerous multiplanetary 
systems which they will detect will provide stimulating
constraints for formation scenarios.

\begin{acknowledgements}
The authors thank the observers of the other HARPS GTO 
sub-programmes who helped measure \object{GJ 581}. We are grateful 
to the staff of La Silla Observatory for their contribution to the 
success of the HARPS project. We are grateful to our colleagues 
Ji-Lin Zhou, Douglas Lin, Su Wang and Katherine Kretke
for having made their study of multiple super-Earth formation 
available in advance of publication. We wish to thank the Programme National 
de Plan\'etologie (INSU-PNP) and the Swiss National Science 
Foundation for their continuous support of our planet-search programs. 
XB acknowledges support from the Funda\c{c}\~ao para a Ci\^encia e 
a Tecnologia (Portugal) in the form of a fellowship (reference
SFRH/BPD/21710/2005) and a program (reference
PTDC/CTE-AST/72685/2006), as well from the Gulbenkian 
Foundation for funding through the ``Programa de Est\'{\i}mulo 
ˆ Investiga\c{c}\~ao''. NCS would like to thank the support from 
Funda\c{c}\~ao para a Ci\^encia e a Tecnologia, Portugal, in the 
form of a grant (references 
POCI/CTE-AST/56453/2004 and PPCDT/CTE-AST/56453/2004), and through program 
Ci\^encia\,2007 (C2007-CAUP-FCT/136/2006).

\end{acknowledgements}

\bibliographystyle{aa}
\bibliography{maBiblio}

\longtab{1}{
\begin{longtable}{cccc}
\caption{\label{tab:rv}
Radial-velocity measurements, associated uncertainties and 
\ion{Ca}{ii} H$+$K index for Gl\,581. All 
radial velocities are relative to the solar system barycenter (and 
not corrected from the 0.20 $\mathrm{m\,s^{-1}}$ perspective
acceleration of GJ~581). Only available
electronically.}\\
%\centering
%\begin{tabular}{c c c c}
\hline\hline
\bf JD-2400000 & \bf RV & \bf Uncertainty & \ion{Ca}{ii} H$+$K\\
 & \bf [km\,s$^{-1}$] & \bf [km\,s$^{-1}$] & \\
\hline
\endfirsthead
\caption{continued.}\\
\hline
\hline
\bf JD-2400000 & \bf RV & \bf Uncertainty & \ion{Ca}{ii} H$+$K\\
 & \bf [km\,s$^{-1}$] & \bf [km\,s$^{-1}$] & \\
\hline
\endhead
\hline
\endfoot
53152.712894&	-9.21879&	 0.0011& 0.751\\
53158.663460&	-9.22730&	 0.0012& 0.847\\
53511.773341&	-9.21549&	 0.0012& 0.885\\
53520.744746&	-9.19866&	 0.0013& 0.764\\
53573.512039&	-9.20874&	 0.0012& 0.936\\
53574.522329&	-9.19922&	 0.0011& 0.766\\
53575.480749&	-9.20466&	 0.0010& 0.790\\
53576.536046&	-9.21536&	 0.0010& 0.775\\
53577.592603&	-9.21927&	 0.0012& 0.742\\
53578.510713&	-9.20889&	 0.0009& 0.812\\
53578.629602&	-9.20628&	 0.0011& 0.708\\
53579.462557&	-9.19585&	 0.0009& 0.891\\
53579.621049&	-9.19414&	 0.0010& 0.800\\
53585.461770&	-9.20243&	 0.0010& 0.771\\
53586.465158&	-9.21196&	 0.0007& 1.049\\
53587.464704&	-9.22556&	 0.0014& 0.833\\
53588.538062&	-9.21678&	 0.0023& 0.736\\
53589.462024&	-9.20322&	 0.0007& 0.819\\
53590.463895&	-9.19562&	 0.0008& 0.916\\
53591.466484&	-9.20138&	 0.0007& 1.099\\
53592.464813&	-9.21424&	 0.0008& 0.898\\
53606.551679&	-9.19397&	 0.0019& 0.961\\
53607.507527&	-9.19849&	 0.0010& 0.733\\
53608.482643&	-9.21273&	 0.0012& 0.771\\
53609.488452&	-9.22225&	 0.0015& 0.670\\
53757.877319&	-9.20435&	 0.0010& 0.627\\
53760.875475&	-9.21109&	 0.0013& 1.057\\
53761.859216&	-9.20177&	 0.0013& 0.870\\
53811.846943&	-9.20333&	 0.0012& 0.889\\
53813.827017&	-9.21874&	 0.0009& 0.832\\
53830.836957&	-9.21013&	 0.0009& 0.751\\
53862.701441&	-9.20862&	 0.0009& 0.674\\
53864.713662&	-9.19447&	 0.0010& 0.581\\
53867.752171&	-9.21747&	 0.0011& 0.550\\
53870.696603&	-9.20299&	 0.0011& 0.730\\
53882.657763&	-9.21956&	 0.0009& 0.744\\
53887.690738&	-9.21332&	 0.0009& 0.736\\
53918.621751&	-9.20079&	 0.0011& 0.637\\
53920.594947&	-9.22608&	 0.0010& 0.696\\
53945.543122&	-9.19820&	 0.0010& 0.673\\
53951.485927&	-9.20254&	 0.0009& 0.655\\
53975.471596&	-9.21567&	 0.0010& 0.895\\
53979.543975&	-9.21606&	 0.0012& 0.693\\
54166.874182&	-9.22284&	 0.0011& 0.577\\
54170.853961&	-9.20059&	 0.0009& 0.704\\
54194.872349&	-9.22170&	 0.0011& 0.681\\
54196.750384&	-9.19345&	 0.0012& 0.620\\
54197.845039&	-9.19264&	 0.0012& 0.631\\
54198.855509&	-9.21129&	 0.0012& 0.662\\
54199.732869&	-9.21393&	 0.0010& 0.756\\
54200.910915&	-9.20829&	 0.0010& 0.841\\
54201.868554&	-9.19897&	 0.0010& 0.763\\
54202.882595&	-9.19475&	 0.0010& 0.823\\
54228.741561&	-9.19964&	 0.0011& 0.799\\
54229.700478&	-9.19722&	 0.0014& 0.736\\
54230.762136&	-9.20982&	 0.0010& 0.612\\
54234.645916&	-9.19421&	 0.0011& 0.532\\
54253.633168&	-9.21733&	 0.0010& 0.605\\
54254.664810&	-9.21238&	 0.0010& 0.569\\
54291.568850&	-9.21520&	 0.0013& 0.615\\
54292.590814&	-9.20807&	 0.0009& 0.558\\
54293.625868&	-9.19804&	 0.0010& 0.575\\
54295.639448&	-9.21852&	 0.0011& 0.490\\
54296.606114&	-9.22915&	 0.0012& 0.515\\
54297.641939&	-9.21556&	 0.0010& 0.647\\
54298.567600&	-9.20023&	 0.0011& 0.653\\
54299.622195&	-9.19742&	 0.0015& 0.647\\
54300.619110&	-9.20792&	 0.0010& 0.700\\
54315.507494&	-9.19471&	 0.0016& 0.501\\
54317.480847&	-9.21647&	 0.0010& 0.684\\
54319.490527&	-9.20583&	 0.0013& 0.711\\
54320.544072&	-9.19620&	 0.0010& 0.733\\
54323.507050&	-9.21928&	 0.0032& 0.674\\
54340.555781&	-9.20957&	 0.0009& 0.695\\
54342.486204&	-9.18958&	 0.0010& 0.592\\
54349.515163&	-9.21982&	 0.0010& 0.680\\
54530.855660&	-9.20187&	 0.0010& 0.587\\
54550.831274&	-9.20188&	 0.0009& 0.600\\
54553.803722&	-9.22133&	 0.0008& 0.666\\
54563.838002&	-9.21136&	 0.0009& 0.674\\
54566.761147&	-9.20787&	 0.0011& 0.695\\
54567.791671&	-9.19972&	 0.0010& 0.592\\
54569.793302&	-9.22225&	 0.0010& 0.680\\
54570.804248&	-9.22375&	 0.0010& 0.587\\
54571.818376&	-9.20449&	 0.0011& 0.600\\
54587.861968&	-9.20505&	 0.0015& 0.666\\
54588.838799&	-9.19985&	 0.0011& 0.637\\
54589.827493&	-9.20286&	 0.0011& 0.655\\
54590.819634&	-9.21398&	 0.0010& 0.745\\
54591.817120&	-9.22848&	 0.0015& 0.840\\
54592.827337&	-9.21685&	 0.0009& 0.697\\
54610.742932&	-9.18919&	 0.0011& 0.811\\
54611.713477&	-9.19994&	 0.0009& 0.804\\
54616.713028&	-9.19947&	 0.0013& 0.713\\
54639.686739&	-9.21881&	 0.0011& 0.611\\
54640.654409&	-9.21858&	 0.0013& 0.594\\
54641.631706&	-9.20599&	 0.0010& 0.614\\
54643.644998&	-9.20581&	 0.0013& 0.605\\
54644.587028&	-9.21830&	 0.0012& 0.575\\
54646.625357&	-9.21515&	 0.0012& 0.636\\
54647.579119&	-9.20087&	 0.0011& 0.487\\
54648.484817&	-9.19842&	 0.0011& 0.568\\
54661.553706&	-9.22001&	 0.0012& 0.728\\
54662.549408&	-9.20873&	 0.0013& 0.763\\
54663.544866&	-9.19438&	 0.0011& 0.683\\
54664.553042&	-9.19450&	 0.0012& 0.759\\
54665.569376&	-9.20207&	 0.0010& 0.712\\
54672.531723&	-9.21359&	 0.0018& 0.809\\
54674.524120&	-9.19858&	 0.0013& 0.645\\
54677.505114&	-9.21696&	 0.0011& 0.713\\
54678.556785&	-9.20646&	 0.0011& 0.748\\
54679.504034&	-9.19434&	 0.0015& 0.770\\
54681.514143&	-9.20461&	 0.0014& 0.781\\
54682.503343&	-9.21537&	 0.0013& 0.877\\
54701.485074&	-9.19390&	 0.0012& 0.682\\
54703.513042&	-9.21009&	 0.0012& 0.751\\
54708.479050&	-9.21261&	 0.0012& 0.621\\
54721.473033&	-9.22361&	 0.0012& 0.658\\
54722.472371&	-9.20543&	 0.0011& 0.684\\
\hline
%\end{tabular}
\end{longtable}
}

\end{document}